\documentclass[letter]{aa}
\usepackage{graphicx}
\usepackage{color}
\usepackage[varg]{txfonts}
\graphicspath{{./Figures/}}

\begin{document}

\title{A "no-drift" runaway pile-up of pebbles in protoplanetary disks in which midplane turbulence increases with radius}
\titlerunning{No-drift runaway pile-up}

   \author{Ryuki Hyodo
          \inst{1}
          \and
          Shigeru Ida
          \inst{2}
          \and
          Tristan Guillot
          \inst{3}
           }
         \institute{ISAS/JAXA, Sagamihara, Kanagawa, Japan (\email{hyodo@elsi.jp})
         \and Earth-Life Science Institute, Tokyo Institute of Technology, Meguro-ku, Tokyo 152-8550, Japan 
         \and Laboratoire J.-L.\ Lagrange, Universit\'e C\^ote d'Azur, Observatoire de la C\^ote d'Azur, CNRS, F-06304 Nice, France}

\date{DRAFT:  \today}

\abstract
{A notable challenge of planet formation is to find a path to directly form planetesimals from small particles.}
{We aim to understand how drifting pebbles pile up in a protoplanetary disk with a nonuniform turbulence structure.}
{We consider a disk structure in which the midplane turbulence viscosity increases with the radius in protoplanetary disks, such as in the outer region of a dead zone. We perform 1D diffusion-advection simulations of pebbles that include back-reaction (the inertia) to the radial drift and the vertical and radial diffusions of pebbles for a given pebble-to-gas mass flux.}
{We report a new mechanism, the "no-drift" runaway pile-up, that leads to a runaway accumulation of pebbles in disks, thus favoring the formation of planetesimals by streaming and/or gravitational instabilities. This occurs when pebbles drifting in from the outer disk and entering a dead zone experience a decrease in vertical turbulence. The scale height of the pebble subdisk then decreases, and, for small enough values of the turbulence in the dead zone and high values of the pebble-to-gas flux ratio, the back-reaction of pebbles on gas leads to a significant decrease in their drift velocity and thus their progressive accumulation. This occurs when the ratio of the flux of pebbles to that of the gas is large enough that the effect dominates over any Kelvin-Helmholtz shear instability. This process is independent of the existence of a pressure bump.}
{}

\authorrunning{R. Hyodo, S. Ida, T. Guillot}
\maketitle 

\begin{figure*}[h]
        \centering
        \resizebox{0.9\hsize}{!}{ \includegraphics{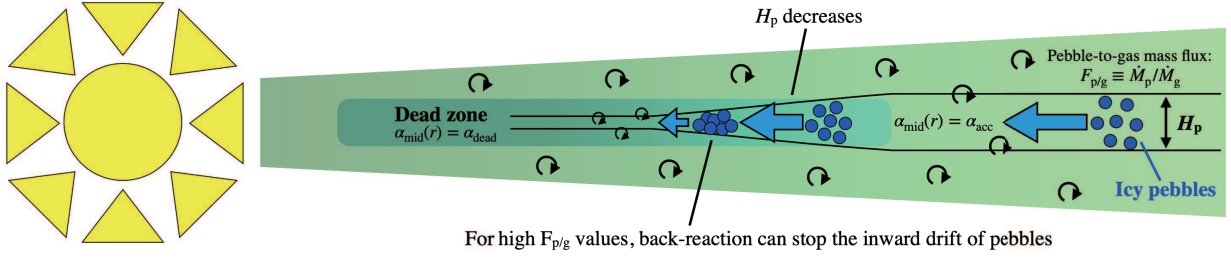} }
        \caption{Schematic illustration of pebble drift and its pile-up within a protoplanetary disk with a dead zone. The disk gas accretion is characterized by $\alpha$-parameter $a_{\rm acc}$, while the midplane diffusivity, being a dead zone ($\alpha_{\rm dead} \ll \alpha_{\rm acc}$) in the inner region, is characterized by $\alpha_{\rm mid}$. During the inward drift of pebbles, the pebble scale height $H_{\rm p}$ decreases as $H_{\rm p} \propto \alpha_{\rm mid}^{1/2}$ until a KH instability prevents it from becoming smaller. A smaller $H_{\rm p}$ leads to an elevated local midplane concentration of pebbles within a thinner midplane layer. The elevated midplane pebble-to-gas ratio causes the back-reaction to be more effective in reducing the radial drift velocity of pebbles. Such a physical interplay with a sufficiently large pebble-to-gas mass flux results in a progressive accumulation of pebbles in a runaway fashion (i.e., the "no-drift" runaway pile-up).}
\label{fig_summary}
\end{figure*}

\section{Introduction} \label{sec_intro}
Forming planetesimals directly from small particles in an evolving protoplanetary disk is a major challenge in planet formation due to the "growth barrier" \citep{Blu00,Zso10} and the "drift barrier" \citep{Whi72,Wei77}. Streaming instability (SI) may be a promising mechanism for forming planetesimals directly from pebbles, although it requires special conditions: the Stokes number $\tau_{\rm s} \gtrsim 0.01$ and a local elevated solid-to-gas ratio \citep[$Z \equiv \rho_{\rm p}/\rho_{\rm g}  \gtrsim 1$, where $\rho_{\rm p}$ and $\rho_{\rm g}$ are the spatial densities of pebbles and gas, respectively;][]{You04,You05,Car15}. How, when, and whether such conditions are met in an evolving protoplanetary disk are still a matter of intense debate.

Different mechanisms have been proposed for planetesimal formation via the local enhancement of materials. They include the diffusive redistribution and recondensation of water vapor outside the snow line \citep[e.g.,][]{Ste88,Cie06,Ros13}, and detailed numerical simulations have been performed \citep[e.g.,][]{Dra17,Sch17,Hyo19,Gar20,Hyo20}. Other mechanisms include pile-ups at pressure maxima. These pressure maxima can be created by the planet's gravity \citep[e.g.,][]{Dip17,Kan18}, by a sharp change in the local magneto-rotational instability- (MRI-) driven turbulence structure at an evaporation front \citep[e.g.,][]{Kre07,Bra08}, by a change in the gas accretion velocity at the inner \citep[e.g.,][]{Cha14,Ued19,Cha19} and outer \citep[e.g.,][]{Pin16} edges of a dead zone, and/or by the change in the gas profile due to gas-pebble friction (back-reaction) combined with pebble growth \citep{Gon17}. However, a positive pressure gradient may not be likely because the non-magnetohydrodynamic (non-MHD) effects erase a clear transition between dead and active zones \citep[e.g.,][]{Mor17}. The back-reaction of solids onto the gas would smooth out the bump \citep[e.g.,][]{Tak16,Kan18}. Further studies are required to assess these likelihoods in an evolving disk.

Over the last few decades, MHD simulations have shown that a region where ionization is too low for the MRI to operate (i.e., a "dead zone"; \cite{Gam96}) might ubiquitously exist in the inner part of the disk midplane, and that only surface layers are magnetically active, supporting accretion \citep[Fig.~\ref{fig_summary};][]{Gre15,Sim15,Bai13,Bai16,Mor17}. The transition between active and dead zones might be smooth (i.e., the $\alpha$-parameter in the disk midplane $\alpha_{\rm mid}$ increases as a function of the distance to the star $r$), and $\alpha_{\rm dead}=10^{-5} - 10^{-3}$ is reported for the $\alpha$-parameter within a dead zone \citep[e.g.,][]{Gre15,Sim15,Mor19}. We note that the sound waves propagated from MRI-active surface layers \citep[e.g.,][]{Oku11,Yan18} and the vertical shear instability \citep[VSI; e.g.,][]{Sto16,Flo20} could induce vertical mixing with an equivalent $\alpha$ for vertical diffusivity up to $\sim 10^{-3}$, even within a dead zone. These results imply nonuniform disk turbulence structures. 

In this letter, we consider a toy model of a protoplanetary disk with a dead zone located at the inner region of the disk. We consider different $\alpha$-parameters for the gas accretion and for the pebble motion. Including the effects of the back-reaction of pebbles onto gas, which slows the drift velocity of pebbles, we study how drifting pebbles within a dead zone pile up. We demonstrate that a runaway pile-up of pebbles occurs for a sufficiently large pebble-to-gas mass flux $F_{\rm p/g}$ when pebbles reach a critical low-level of turbulence; we call this the "no-drift" (ND) runaway pile-up (the summary of this mechanism is shown by a schematic illustration in Fig.~\ref{fig_summary}). It is worth mentioning that the newly reported ND mechanism does not require the snow line or the pressure maxima.

In Section \ref{sec_model}, we describe the disk models and settings of our 1D simulations that solve the diffusion-advection of drifting pebbles. In Section \ref{sec_results}, we show the results of our 1D simulations as well as analytical arguments. In Section \ref{sec_summary}, we summarize this letter.

\section{Models and settings} \label{sec_model}

\subsection{Gas structure}
Here, we adopt  a toy model where the gas accretion toward the central star is characterized by a non-dimensional $\alpha$-parameter $\alpha_{\rm acc}$ (Fig.~\ref{fig_summary}). Using the prescription of the classical $\alpha$-accretion disk model \citep{Sha73,Lyn74}, the surface density of the gas is given as 
\begin{equation}
        \Sigma_{\rm g} = \frac{\dot{M}_{\rm g}}{3\pi \nu_{\rm acc}} = \frac{\dot{M}_{\rm g}}{3\pi \alpha_{\rm acc}c_{\rm s}^2 \Omega_{\rm K}^{-1}},
\label{eq_sigma_g}
\end{equation}
where $\dot{M}_{\rm g}$ and $\nu_{\rm acc}=\alpha_{\rm acc}c_{\rm s}^2 \Omega_{\rm K}^{-1}$ are the gas mass accretion rate and the effective viscosity, respectively ($c_{\rm s}$ is the gas sound velocity and $\Omega_{\rm K}$ is the Keplerian orbital frequency). 

The gas rotates at sub-Keplerian speed. The degree of the deviation of the gas rotation frequency from that of Keplerian $\eta$ is given by
\begin{equation}
        \eta \equiv \frac{\Omega_{\rm K} - \Omega}{\Omega_{\rm K}} = -\frac{1}{2} \frac{\partial \ln P_{\rm g}}{\partial \ln r} \left( \frac{H_{\rm g}}{r} \right)^2 = C_{\rm \eta} \left( \frac{H_{\rm g}}{r} \right)^2 ,
\end{equation}
where $\Omega$, $P_{\rm g}$, and $H_{\rm g}$ are the gas orbital frequency, the gas pressure, and the gas scale height, respectively, and $C_{\rm \eta}$ is defined as  
\begin{equation}
        C_{\rm \eta} \equiv -\frac{1}{2} \frac{\partial \ln P_{\rm g}}{\partial \ln r} ,
\end{equation}
which depends on the temperature profile (e.g., $C_{\rm \eta}=11/8$ for $T \propto r^{-1/2}$). 

Using the $\alpha$-disk prescription, the gas accretion velocity $v_{\rm g}$ is written as  
\begin{align}
        v_{\rm g} & = -\frac{3\nu_{\rm acc}}{2r} = -\frac{3\alpha_{\rm acc} H_{\rm g}^{2} \Omega_{\rm K}}{2 r} = -\frac{3\alpha_{\rm acc}}{2} \left( \frac{H_{\rm g}}{r} \right)^{2} v_{\rm K} ,\\ 
        & \simeq -\frac{3}{2} \alpha_{\rm acc} \eta  v_{\rm K} \left( -\frac{1}{2} \frac{d \ln P_{\rm g}}{d \ln r} \right)^{-1} = -\frac{3}{2} \alpha_{\rm acc} \eta  v_{\rm K} C_{\rm \eta}^{-1} ,
\label{eq_v_gas}
\end{align}
where a negative sign indicates accretion toward the central star.

\subsection{Pebbles in the disk midplane}

The radial drift velocity of pebbles, including the effects of gas-solid friction $-$ drift back-reaction (hereafter Drift-BKR) $-$ is given as \citep{Ida16,Sch17,Hyo19}
\begin{align}
\label{eq_vp}
        v_{\rm p} = - \frac{\Lambda}{1+\Lambda^{2}\tau_{\rm s}^{2}} \left( 2\tau_{\rm s}\Lambda \eta v_{\rm K} - v_{\rm g} \right) ,
\end{align}
where $\tau_{\rm s}$ is the Stokes number of pebbles. $\Lambda \equiv \rho_{\rm g}/(\rho_{\rm g} + \rho_{\rm p}) = 1/(1+Z)$ characterizes the strength of the back-reaction due to the pile-up of pebbles, where $Z \equiv \rho_{\rm p}/\rho_{\rm g}$ is the midplane pebble-to-gas density ratio ($\rho_{\rm g}=\Sigma_{\rm g}/\sqrt{2\pi}H_{\rm g}$ and $\rho_{\rm p}=\Sigma_{\rm p}/\sqrt{2\pi} H_{\rm p}$, where $\Sigma_{\rm p}$ and $H_{\rm p}$ are the surface density and scale height of pebbles, respectively). 

The disk midplane could be too weakly ionized for the MRI to operate (the dead zone with extremely weak turbulence; \cite{Gam96}). Thus, the disk midplane could have very different turbulence from that of gas accretion, which controls the disk surface density (i.e., $\alpha_{\rm acc}$). In this letter, we use an effective viscous parameter $\alpha_{\rm mid}$ to characterize the radial and vertical diffusion processes of pebbles near the disk midplane (Fig.~\ref{fig_summary}).

The scale height of pebbles characterizes the degree of pebble concentration in the disk midplane ($\rho_{\rm p} \propto H^{-1}_{\rm p}$). In the steady state, the scale height of pebbles is regulated by the vertical turbulent stirring \citep{Dub95,You07,Oku12,Hyo19} as
\begin{align}
\label{eq_Hp_tur}
        H_{\rm p,tur} = \left( 1 + \frac{\tau_{\rm s}}{\alpha_{\rm mid} \left( 1+ Z \right)^{-K} } \right)^{-1/2} H_{\rm g},
\end{align}
where a coefficient $K$ characterizes the strength of the back-reaction onto the diffusivity \citep{Hyo19,Ida20} and $K=1$ is used (the choice of the $K$ value does not alter the conclusion -- see Appendix \ref{sec_app_K0} for the $K=0$ case).
 
For a small $\alpha_{\rm mid}$ (i.e., for a small $H_{\rm p,tur}$), a vertical shear Kelvin-Helmholtz (KH) instability prevents a further decrease in the pebble scale height. This minimum scale height, $H_{\rm p, KH}$, is \citep{Hyo20} 
\begin{align}
        H_{\rm p,KH} \simeq Ri^{1/2} \frac{Z^{1/2}}{ \left( 1+Z \right)^{3/2} } C_{\rm \eta} \left( \frac{H_{\rm g}}{r} \right) H_{\rm g} = Ri^{1/2} \frac{Z^{1/2}}{ \left( 1+Z \right)^{3/2} } \eta r,
\label{eq_Hp_KH}
\end{align}
which is valid for $Z \lesssim 1$, and $Z \gg 1$ indicates a gravitational collapse. Thus, the scale height of pebbles $H_{\rm p}$ is given as
\begin{align}
\label{eq_Hp}
        H_{\rm p} = \max \left\{ H_{\rm p,tur}, H_{\rm p,KH} \right\} .
\end{align}
%

\begin{figure*}[h]
        \centering
        \resizebox{0.9\hsize}{!}{ \includegraphics{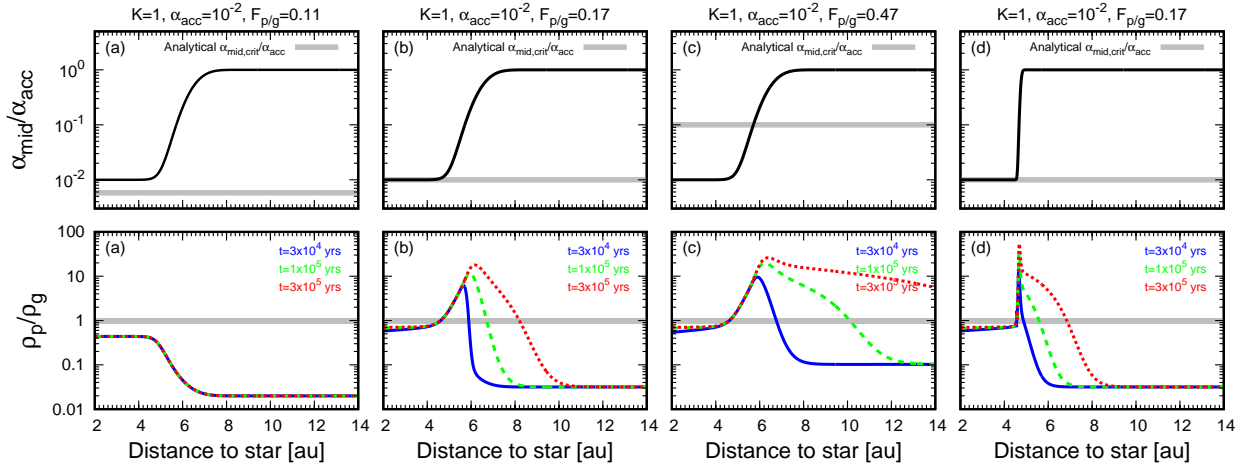} }
        \caption{$\alpha_{\rm mid}/\alpha_{\rm acc}$ (top panels) and $Z=\rho_{\rm p}/\rho_{\rm g}$ (bottom panels) as a function of the distance to a star in the 1D numerical simulations. Included are the cases where $\alpha_{\rm acc}=10^{-2}$ and $\tau_{\rm p}=0.1$. From left to right, the panels show the cases where $(F_{\rm p/g}, \alpha_{\rm dead}, r^{*}, \Delta r_{\rm tra})$ = $(0.11, 10^{-4}, 9{\rm \, au}, 5{\rm \, au})$, $(0.17,10^{-4},9{\rm \, au},5{\rm \, au})$, $(0.47,10^{-4},9{\rm \, au},5{\rm \, au})$, and $(0.17,10^{-4},5{\rm \, au}, 0.5{\rm \, au})$. The gray lines in the top panels represent the analytical critical $\alpha_{\rm mid}/\alpha_{\rm acc}$ below which the ND runaway pile-up is expected to occur for a given $F_{\rm p/g}$ (Eq.~(\ref{eq_ND_cri}) with $F_{\rm p/g} > F_{\rm p/g,crit2}$). The gray lines in the bottom panels show the critical $Z$ (i.e., $Z_{\rm crit}=1$) above which the ND runaway pile-up occurs (Section \ref{sec_analytical}). The blue, green, and red lines in the bottom panels are those at $t=3 \times 10^4$ yrs, $t=1 \times 10^5$ yrs, and $t=3 \times 10^5$ yrs from the beginning of the calculations, respectively. Panel (a) shows the system that reaches a steady-state, while the other cases (panels (b), (c), and (d)) show the ND runaway pile-ups.}
\label{fig_results}
\end{figure*}

\subsection{Numerical settings} \label{sec_settings}
We performed 1D diffusion-advection simulations that included the back-reaction (the inertia) to radial drift of pebbles that slows the pebble drift velocity as pile-up proceeds \citep[see details in][]{Hyo19,Hyo20}. The drift velocity of pebbles is given by Eq.~(\ref{eq_vp}), while the diffusivity is given as $D_{\rm p} = \alpha_{\rm mid}c_{\rm s}^{2} \Omega_{\rm K}^{-1} \Lambda^{K} / (1+{\rm \tau_{\rm s}^2})$. We included back-reaction onto the diffusivity of pebbles (with $K=1$). We assumed that pebbles drift outside the snow line, and the sublimation of pebbles was neglected (i.e., the dead zone exists beyond the snow line). The Stokes number of pebbles was set to be constant: $\tau_{\rm s}=0.1$ \citep{Oku16,Ida16}. The disk midplane temperature was set to $T(r) = 150 {\, \rm K} \times ( r/3 {\, \rm au} )^{-\beta}$ ($\beta=1/2$). The surface density of the gas (molecular weight of $\mu_{\rm g}=2.34$) is described by $\Sigma_{\rm g}=\dot{M}_{\rm g}/3 \pi \nu_{\rm g}$, where $\nu_{\rm g} = \alpha_{\rm acc} c_{\rm s}^2 \Omega_{\rm K}^{-1}$. We used $\alpha_{\rm acc}=10^{-2}$ and $\dot{M}_{\rm g}=10^{-8}$ $M_{\odot}$/year. These led to $C_{\rm \eta} = 11/8$ for $\Sigma_{\rm g} \propto r^{-1}$ and $T \propto r^{-1/2}$.

We considered a dead zone in the inner region of the disk midplane, and we used a non-dimensional turbulence parameter in the midplane $\alpha_{\rm mid}$, which differs from the one that characterizes gas accretion, $\alpha_{\rm acc}$ (Eq.~(\ref{eq_sigma_g})). The midplane $\alpha_{\rm mid}$ is modeled as 
\begin{equation}
        \alpha_{\rm mid}(r) = \alpha_{\rm dead} +  \left( \frac{\alpha_{\rm acc} - \alpha_{\rm dead}}{2} \right) \left[ \mathrm{erf} \left( 3 + \frac{6\left( r - r^{*} \right) }{\Delta r_{\rm tra}} \right) + 1 \right],
\label{eq_alpha_mid}
\end{equation}
where: $\alpha_{\rm acc}$ and $\alpha_{\rm dead}$ are those outside and inside a dead zone; $r^{*}$ is the innermost radial distance, where $\alpha_{\rm mid}=\alpha_{\rm acc}$; and $\Delta r_{\rm tra}$ is the radial width of the transition from $\alpha_{\rm mid}=\alpha_{\rm acc}$ to $\alpha_{\rm mid}=\alpha_{\rm dead}$. As shown below, the ND mode occurs for an arbitrary choice of the dead zone structure, that is, irrespective of a sharp or a smooth change between the active and dead zones (i.e., an arbitrary choice of $\alpha_{\rm dead}$, $\alpha_{\rm acc}$, $r^{*}$, and $\Delta r_{\rm tra}$), as long as it satisfies that $\alpha_{\rm mid}$ is smaller than a threshold value (Section \ref{sec_critical}).
 
At the beginning of the 1D simulations, we set the pebble-to-gas mass flux $F_{\rm p/g}$ at the outer boundary ($r_{\rm out}=15$ au), and $F_{\rm p/g}$ at $r_{\rm out}$ was fixed throughout the simulations. We note that \cite{Elb20} performed a time-dependent simulation of disk formation to find a high $F_{\rm p/g}$ variation between $\mathcal{O}(10^{-4})$ to $\mathcal{O}(1)$ in a single disk evolution \citep[see also][]{Ida20}.

\section{Results}
\label{sec_results}

\subsection{Numerical results} \label{sec_numerical}
Figure \ref{fig_results} shows the results of our 1D simulations for different combinations of $F_{\rm p/g}$ and the midplane turbulence structures (i.e., different choices of $\alpha_{\rm dead}$, $r^{*}$, and $\Delta r_{\rm tra}$ in Eq.~(\ref{eq_alpha_mid})) for the case of $\alpha_{\rm acc}=10^{-2}$. The top panels show given turbulence structures (i.e., $\alpha_{\rm mid}(r)/\alpha_{\rm acc}$) and the bottom panels show the resultant midplane pebble-to-gas ratio in their spatial densities (i.e., $\rho_{\rm p}/\rho_{\rm g}$). The gray lines in the top panels are the analytically derived critical $\alpha_{\rm mid}/\alpha_{\rm acc}$ below which a runaway pile-up of pebbles is expected to occur for a given $F_{\rm p/g}$ (Eq.~(\ref{eq_ND_cri}) with $\tau_{\rm s}=0.1$).

For a small $F_{\rm p/g}$ (e.g., $F_{\rm p/g}=0.11$; panel (a) in Fig.~\ref{fig_results}), the system reaches a steady state. For this given $F_{\rm p/g}$, analytical arguments (the gray line) predict the requirement of a smaller $\alpha_{\rm mid}/\alpha_{\rm acc}$ for a runaway pile-up of pebbles to occur (i.e., the black line in Fig.~\ref{fig_results} needs to be below the gray line). 

For a larger $F_{\rm p/g}$ (e.g., $F_{\rm p/g}=0.17$; panel (b) in Fig.~\ref{fig_results}) that has the same $\alpha_{\rm mid}/\alpha_{\rm acc}$ structure as that of panel (a), pebbles progressively pile up over time. Continuous pile-up occurs as the drift velocity of pebbles progressively decreases due to strong Drift-BKR and as the pile-up efficiently proceeds. For the $F_{\rm p/g}$ given here, the gray line overlaps with the black line (i.e., the analytical requirement for the runaway pile-up is met) and the analytical arguments (Section \ref{sec_analytical}) show a very good accordance with 1D numerical simulations. For an even larger $F_{\rm p/g}$ with the same $\alpha_{\rm mid}(r)/\alpha_{\rm acc}$ as those in panels (a) and (b), a runaway pile-up of pebbles occurs more quickly and significantly (panel (c) in Fig.~\ref{fig_results}).

Comparing panels (b) and (d) in Fig.~\ref{fig_results} (the same $F_{\rm p/g}$, $\alpha_{\rm dead}$, and $\alpha_{\rm acc}$ but different shapes and locations of the transition between active and dead zones), 1D simulations demonstrate that a runaway pile-up of pebbles occurs irrespective of the shape and location of the dead zone as long as $F_{\rm p/g}$ is sufficiently large. This shows the robustness of this new mechanism of runaway pebble pile-up.

Numerical results show that when the change in the midplane turbulence on $r$ is smooth (i.e., $\alpha_{\rm mid}$), the pile-up of pebbles efficiently propagates toward the outer region in the disk (compare panels (b) and (d) in Fig.~\ref{fig_results}). This indicates that, when the ND runaway pile-up occurs at a given radial distance, a progressive pile-up of pebbles could occur even at a greater radial distance where the ND conditions are not met, leading to a global pile-up of pebbles for the SI to operate. 

When $\alpha_{\rm acc}=10^{-3}$, 1D simulations only show a runaway pile-up of pebbles for a very large value of $F_{\rm p/g}$ ($F_{\rm p/g} \gtrsim 0.8$). This is because the surface density of the gas increases (Eq.~(\ref{eq_sigma_g})) for a smaller $\alpha_{\rm acc}$, and $\rho_{\rm p}/\rho_{\rm g}$ correspondingly decreases.

\subsection{Analytical arguments}
\label{sec_analytical}

\begin{figure*}[h]
        \centering
        \resizebox{0.7\hsize}{!}{ \includegraphics{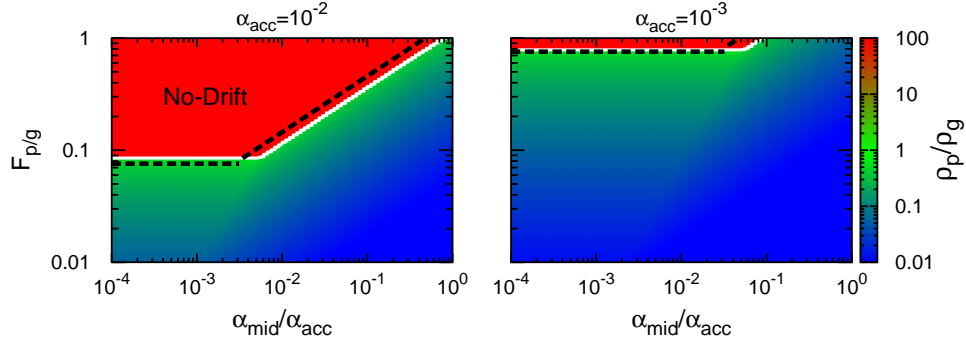} }
        \caption{Midplane pebble-to-gas ratio $Z \equiv \rho_{\rm p}/\rho_{\rm g}$ in the $\alpha_{\rm mid}/\alpha_{\rm acc} - F_{\rm p/g}$ space. Included are the cases where $r=5$ au, $C_{\rm \eta}=11/8$ for $T \propto r^{-1/2}$, $\tau_{\rm s}=0.1$, and $Ri=0.5$. The left and right panels are the cases where $\alpha_{\rm acc}=10^{-2}$ and $\alpha_{\rm acc}=10^{-3}$, respectively. The color contours are obtained by directly solving Eq.~(\ref{eq_peb_pileup}). The red regions indicate the no-drift runaway pile-up. The diagonal dashed black line (Eq.~(\ref{eq_Fpg_cri_regime1})) and the horizontal dashed black line (Eq.~(\ref{eq_Fpg_cri_regime2})) are the analytically derived critical $F_{\rm p/g,crit}$ above which the ND runaway pile-up takes place. The critical $Z$ above which the ND runaway pile-up occurs (i.e., $Z_{\rm crit}=1$) is shown by the white lines. The parameter map is very weakly dependent on the radial distance to the star (see Eqs.~(\ref{eq_Fpg_cri_regime1}) and (\ref{eq_Fpg_cri_regime2})), and the ND mode can occur irrespective of the shape and position of a dead zone.}
\label{fig_ND_map}
\end{figure*}

\subsubsection{Midplane concentration of pebbles}
Below, we discuss the concentration of pebbles in the disk midplane considering Drift-BKR. Our arguments are made using the analytical equation~presented below (Eq. (\ref{eq_peb_pileup})). We solve this equation using two distinct approaches: a direct numerical approach and an analytical approach with some approximation. In both ways, we identify a parameter space ($F_{\rm p/g} - \alpha_{\rm acc} - \alpha_{\rm mid}$ space) where the ND runaway pile-up of drifting pebbles could take place.  

The concentration of pebbles at the midplane is written as
\begin{equation}
        Z \equiv \frac{\rho_{\rm p}}{\rho_{\rm g}} = \frac{\Sigma_{\rm p}}{\Sigma_{\rm g}} h_{\rm p/g}^{-1} = \frac{v_{\rm g}}{v_{\rm p}} h_{\rm p/g}^{-1} F_{\rm p/g} ,
\label{eq_peb_pileup}
\end{equation}
where $h_{\rm p/g} \equiv H_{\rm p}/H_{\rm g}$ and $v_{\rm p}$ are functions of $\Lambda(Z)$. We directly find a solution (i.e., solve for $Z$) for Eq.~(\ref{eq_peb_pileup}) for a given $F_{\rm p/g}$ and $\alpha_{\rm mid}$. The color contours in Fig.~\ref{fig_ND_map} show directly obtained $\rho_{\rm p}/\rho_{\rm g}$ and the red region is where no steady-state solution is found (i.e., the radial drift of pebbles is halted due to Drift-BKR). The dashed lines in Fig.~\ref{fig_ND_map} are analytically derived critical $F_{\rm p/g,crit}$, above which no steady-state solution is found for $\rho_{\rm p}/\rho_{\rm g}$ (i.e., the ND runaway pile-up). The analytical estimations reproduce the direct solutions well.

The $F_{\rm p/g,crit}$ can be divided into two different regimes.
In the first regime, $H_{\rm p,tur} > H_{\rm p,KH,max}$, where $H_{\rm p,KH,max}$ is the maximum scale height of pebbles regulated by a KH instability. In this case, using Eqs.~($\ref{eq_v_gas}$) and (\ref{eq_vp}), the vertically averaged metallicity of pebbles is given as
\begin{align}
\label{eq_Z_peb}
        Z_{\rm \Sigma}  & \equiv  \frac{\Sigma_{\rm p}}{\Sigma_{\rm g}} = F_{\rm p/g} \times \frac{v_{\rm g}}{v_{\rm p}} ,\\
          & \simeq  F_{\rm p/g} \times \frac{ -\frac{3}{2} \alpha_{\rm acc} \eta v_{\rm K} C_{\rm \eta}^{-1} } 
          { -  \left(\frac{\Lambda}{1+\Lambda^{2}\tau_{\rm s}^{2}} \right) \left( 2\Lambda \tau_{\rm s} + \frac{3\alpha_{\rm acc}}{2C_{\rm \eta}}  \right) \eta v_{\rm K}},\\
        & \simeq \frac{3 \alpha_{\rm acc} F_{\rm p/g}}{4 \tau_{\rm s} C_{\rm \eta} \Lambda^{2}} = \frac{3 \alpha_{\rm acc} F_{\rm p/g}}{4 \tau_{\rm s} C_{\rm \eta}} \left( 1 + h_{\rm p/g}^{-1} Z_{\rm \Sigma} \right)^{2} ,
\end{align}
where an approximation is made for $\alpha_{\rm acc} \ll \tau_{\rm s} \ll 1$, $Z \ll 1$, $\Lambda \simeq 1$. Solving this equation gives
\begin{equation}
        Z_{\rm \Sigma} = \frac{\left(1 - 2ab \right) \pm \sqrt{1-4ab}}{2ab^{2}} ,
\end{equation}
where $a=\frac{3\alpha_{\rm acc}F_{\rm p/g} }{4\tau_{\rm s} C_{\rm \eta}}$ and $b=h_{\rm p/g}^{-1}$. The $Z_{\rm \Sigma}$ has a real solution when $1-4ab > 0$ and a critical $F_{\rm p/g,crit1}$ for the first regime is given as
\begin{align}
\label{eq_Fpg_cri_regime1}
         F_{\rm p/g, crit1}  & = \frac{C_{\rm \eta}}{3} \frac{\tau_{\rm s}}{\alpha_{\rm acc}} h_{\rm p/g}
          \simeq \frac{ \left( \alpha_{\rm mid}\tau_{\rm s} \right)^{1/2} C_{\rm \eta}}{3\alpha_{\rm acc}} ,\\
         & \simeq 0.15 \times \left( \frac{\alpha_{\rm acc}}{10^{-2}}\right)^{-1} \left( \frac{\alpha_{\rm mid}}{10^{-4}} \right)^{1/2} \left( \frac{\tau_{\rm s}}{0.1} \right)^{1/2}  \left( \frac{C_{\rm \eta}}{11/8} \right) \\
         & \hspace{11em}  \left[  {\rm  for \,} H_{\rm p,tur} > H_{\rm p,KH,max} \right] , \nonumber
\end{align}
where $h_{\rm p/g} \simeq \left( \tau_{\rm s}/\left( \alpha_{\rm mid} (1+Z)^{-K} \right) \right)^{-1/2} \simeq \left( \tau_{\rm s}/\alpha_{\rm mid} \right)^{-1/2}$ for $Z \ll 1$ and $\alpha_{\rm mid} \ll 1+\tau_{\rm s}$ (Eq.~(\ref{eq_Hp})). The above criteria for $F_{\rm p/g, crit1}$ are independent of $r$, and thus the ND mechanism occurs irrespective of the position and shape of a dead zone as long as it satisfies its criteria (i.e., Eq.~(\ref{eq_Fpg_cri_regime1})). The diagonal dashed black line in Fig.~\ref{fig_ND_map} shows Eq.~(\ref{eq_Fpg_cri_regime1}) and is in accordance with the direct solutions of Eq.~(\ref{eq_peb_pileup}) (color contours in Fig.~\ref{fig_ND_map}). The critical $Z$ above which the ND occurs, $Z_{\rm crit}$, is obtained by inserting Eq.~(\ref{eq_Fpg_cri_regime1}) into Eq.~(\ref{eq_peb_pileup}), and $Z_{\rm crit} \sim 1$ for the first regime. Analytical $Z_{\rm crit}=1$ agrees with the direct solution (color contours in Fig.~\ref{fig_ND_map}). We note that because $F_{\rm p/g,crit1}$ is proportional to $C_{\rm \eta}$, it increases with a steeper $\Sigma_{\rm g}$ gradient at an outer boundary of a dead zone. However, the ND mode does not need a sharp boundary. If the boundary is very gradual, the change in $C_{\rm \eta}$ is negligible.

The second regime appears when a KH instability plays a role, which corresponds to $H_{\rm p,tur} < H_{\rm p,KH,max}$. Thus, in this case, the scale height of pebbles is described by $H_{\rm p,KH}$, which is independent of $\alpha_{\rm mid}$. As $\rho_{\rm p}/\rho_{\rm g}$ increases from zero, $H_{\rm p,KH}$ initially increases and reaches its maximum $H_{\rm p,KH,max}$ at $Z = \rho_{\rm p}/\rho_{\rm g} = 1/2$. As $\rho_{\rm p}/\rho_{\rm g}$ further increases, $H_{\rm p,KH}$ decreases\footnote{We note that $H_{\rm p,KH}$ is a mean square root of the vertical height of particles but not the truncated height of the distribution.}. By smoothly connecting with the critical value of $Z_{\rm crit}=1$ in the first regime (i.e., $Z_{\rm crit}=1$ as well in the second regime, which agrees with the direct solution; see the color contours in Fig.~\ref{fig_ND_map}), the critical $F_{\rm p/g}$ in the second regime $F_{\rm p/g,crit2}$ adopts $H_{\rm p,KH}^{Z=1}$ ($H_{\rm p,KH}$ with $Z=1$), and $Z_{\rm crit}$ is given as
\begin{align}
         Z_{\rm crit} & = \frac{\rho_{\rm p}}{\rho_{\rm g}} \mid_{H_{\rm p,KH}^{Z=1}}
           = \frac{\Sigma_{\rm p}}{\Sigma_{\rm g}} \left( h_{\rm p/g,KH}^{Z=1} \right)^{-1} ,\\
          & \simeq  \frac{3 \alpha_{\rm acc} F_{\rm p/g,crit2}}{4 \tau_{\rm s} C_{\rm \eta} (\Lambda_{\rm KH}^{Z=1})^2} \left( h_{\rm p/g,KH}^{Z=1} \right)^{-1} ,
\end{align}
where $h_{\rm p/g,KH}^{Z=1} \equiv H_{\rm p,KH}^{Z=1}/H_{\rm g}$ and an approximation is made for $\alpha_{\rm acc} \ll \tau_{\rm s} \ll 1$ and $\Lambda \simeq 1$. The $\Lambda_{\rm KH}^{Z=1}$ is $\Lambda$ using $H_{\rm p,KH}^{Z=1}$. Here, $Z = 1$ and $\Lambda_{\rm KH}^{Z=1} = 1/2$. Thus, $F_{\rm p/g,crit2}$ with $C_{\rm \eta}=11/8$ is given as 
\begin{align}
\label{eq_Fpg_cri_regime2}
         F_{\rm p/g,crit2} & = \frac{C_{\rm \eta}}{3} \left( \frac{\tau_{\rm s}}{\alpha_{\rm acc}} \right) h_{\rm p/g,KH}^{Z=1} = \frac{11}{24} \left( \frac{\tau_{\rm s}}{\alpha_{\rm acc}} \right) h_{\rm p/g,KH}^{Z=1},\\
        &\simeq 0.06 \times \left( \frac{\alpha_{\rm acc}}{10^{-2}}\right)^{-1} \left( \frac{\tau_{\rm s}}{0.1}\right) \left( \frac{Ri}{0.5}\right)^{1/2} \left( \frac{H_{\rm g}/r}{0.04} \right) , \\
          & \hspace{11em}  \left[  {\rm  for \,} H_{\rm p,tur} < H_{\rm p,KH,max} \right]  \nonumber
\end{align}
which is independent of $\alpha_{\rm mid}$. We note that the above criterion for $F_{\rm p/g, crit2}$ has a very weak dependence on $r$ through $H_{\rm g}/r$. Equation (\ref{eq_Fpg_cri_regime2}) is plotted by a horizontal dashed black line in Fig.~\ref{fig_ND_map}, and it shows a good consistency with the direct solution of Eq.~(\ref{eq_peb_pileup}). Including a KH instability (i.e., the second regime and Eq.~(\ref{eq_Fpg_cri_regime2})) prevents the scale height of pebbles from becoming increasingly smaller as $\alpha_{\rm mid}$ decreases, reducing the parameter range of the ND runaway pile-up\footnote{The vertical shear streaming instability \citep[VSSI;][]{Lin20} might affect the pebble scale height when the midplane diffusivity is very small. If the VSSI dominated over a KH instability for the scale height of pebbles, the boundary of the ND regime would be regulated by the VSSI.}.

\subsubsection{Criteria for the ND runaway pile-up} \label{sec_critical}
Summarizing the analytical arguments above, the scale height of pebbles decreases with decreasing $\alpha_{\rm mid}$ until a KH instability plays a role. As pebbles drift inward from the active to the dead zones in the disk midplane, $\alpha_{\rm mid}$ becomes smaller with decreasing distance to the star $r$, while keeping a constant $F_{\rm p/g}$. This implies that the evolutionary path in the $\alpha_{\rm mid}/\alpha_{\rm acc}-F_{\rm p/g}$ space (Fig.~\ref{fig_ND_map}) is a horizontal shift from right to left for a given fixed $F_{\rm p/g}$ as pebbles drift inward. Therefore, for the ND mode to take place, the following two conditions need to be met.

First, $F_{\rm p/g}$ needs to be sufficiently large to satisfy $F_{\rm p/g} \geq F_{\rm p/g,crit2}$. As $F_{\rm p/g,crit2} \propto 1/\alpha_{\rm acc}$ (Eq.~(\ref{eq_Fpg_cri_regime2})), the ND mode is limited to relatively large $F_{\rm p/g}$ values for the $\alpha_{\rm acc}=10^{-3}$ case ($F_{\rm p/g} \gtrsim 0.8$ at $5$ au), while only moderate $F_{\rm p/g} \gtrsim 0.08$ is required in the case of $\alpha_{\rm acc}=10^{-2}$. 

Second, a sufficiently small $\alpha_{\rm mid}/\alpha_{\rm acc}$ is required  in order to have a sufficiently small scale height of pebbles. Rewriting Eq.~(\ref{eq_Fpg_cri_regime1}) gives a critical $\alpha_{\rm mid}/\alpha_{\rm acc}$ below which the ND occurs for a given $F_{\rm p/g}$ as 
\begin{align}
        \frac{\alpha_{\rm mid,crit}}{\alpha_{\rm acc}} &\equiv \left( \frac{3 F_{\rm p/g}}{C_{\rm \eta}} \right)^2 \alpha_{\rm acc} \tau_{\rm s}^{-1} ,\\
        & \simeq 4.76 \times 10^{-3} \left( \frac{F_{\rm p/g}}{0.1} \right)^2 \left( \frac{C_{\rm \eta}}{11/8} \right)^{-2} \left( \frac{\alpha_{\rm acc}}{10^{-2}} \right) \left( \frac{\tau_{\rm s}}{0.1} \right)^{-1},
\label{eq_ND_cri}
\end{align}
which is independent of $r$. These two criteria  are in very good accordance with the 1D simulations (Section \ref{sec_numerical}).

\section{Conclusions} \label{sec_summary}

In this letter, we studied how drifting pebbles pile up in a protoplanetary disk with a nonuniform turbulence structure, including the back-reaction of pebbles onto the gas that slows the drift velocity of pebbles as pile-up proceeds. We considered that gas accretion is regulated by an $\alpha$-parameter that is distinct from that of midplane turbulence. In the disk midplane, the turbulence strength can decrease with decreasing radial distance (e.g., the case where the inner region of the disk is an MRI-inactive dead-zone and the outer region is MRI-active). Thus, drifting pebbles are further concentrated in the midplane as the scale height decreases. 

We demonstrated a new mechanism for a runaway pile-up of pebbles to occur for a moderate $F_{\rm p/g}$ when pebbles reach a critical level of low turbulence in the dead zone via a continuous slowing of their drift velocity due to the back-reaction (the "no-drift" (ND) runaway pile-up). The ND runaway pile-up occurs irrespective of the shape of the dead zone. Our results imply that SI could also occur as a natural consequence of pebble drift in a protoplanetary disk with a dead zone.

The minimum scale height of pebbles in a dead zone is a critical parameter that characterizes the ND mode. Further studies that include detailed physical processes -- such as a KH instability, VSI, vertical shear streaming instability, and the velocity fluctuation by the sound waves propagated from MRI-active surface layers -- are required.

\begin{acknowledgements}
We thank Satoshi Okuzumi and Shoji Mori for helpful comments and fruitful discussion. R.H. acknowledges the financial support of JSPS Grants-in-Aid (JP17J01269, 18K13600). R.H. also acknowledges JAXA's International Top Young program. S.I. acknowledges the financial support (JSPS Kakenhi 15H02065, MEXT Kakenhi 18H05438). T.G. was partially supported by a JSPS Long Term Fellowship at the University of Tokyo.
\end{acknowledgements}

\bibliography{planetesimals}

\appendix
\onecolumn

\section{The case of $K=0$} \label{sec_app_K0}

\begin{figure*}[h]
        \centering
        \resizebox{\hsize}{!}{ \includegraphics{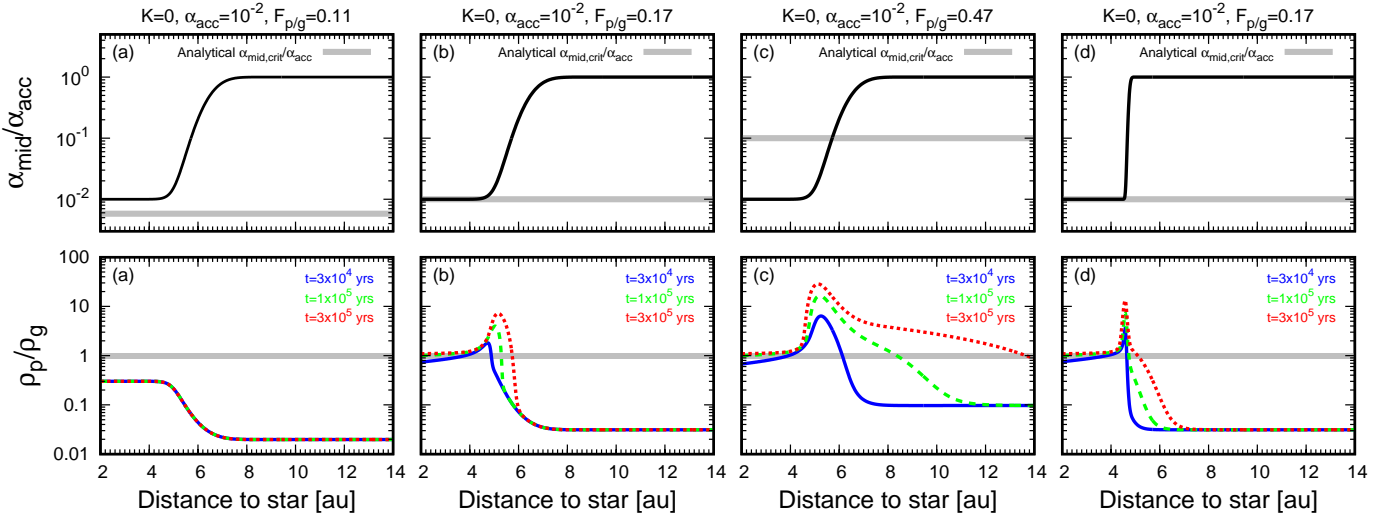} }
        \caption{Same as Fig.~\ref{fig_results} but for the case of $K=0$.}
\label{fig_app_K0}
\end{figure*}

Here, we show the case of $K=0,$ where the diffusivity of pebbles is a fixed constant: $D_{\rm p}=\alpha_{\rm mid}c_{\rm s}^2 \Omega_{\rm K}^{-1}/(1+\tau_{\rm s}^2)$ (i.e., the back-reaction onto the diffusivity is neglected). Figure \ref{fig_app_K0} is the same as Fig.~\ref{fig_results} but for the case of $K=0$. 

As seen in the case of $K=1$ (Fig.~\ref{fig_results}), a runaway pile-up of pebbles within a dead zone occurs as long as $F_{\rm p/g}$ meets the conditions discussed in Section \ref{sec_critical} (the analytical arguments were made for $K=0$). Compared to the case of $K=1$, the radial width and the timescale of the pile-up are in general narrower and longer in the case of $K=0$ because the diffusivity is a constant for $K=0,$ while it becomes smaller for $K=1$ as pile-up proceeds. These additional simulations with $K=0$  further show the robustness of the physical mechanism presented in this letter and demonstrate that Drift-BKR, which slows the drift velocity of pebbles, is a fundamental ingredient for the ND mechanism.

\end{document}